\documentclass[a4paper, 12pt]{article}

\usepackage[dvips]{epsfig}

\usepackage{amsmath}

\usepackage{amssymb}

\begin{document}

\begin{flushright}

{\sc BI-TP 2004/12}

\end{flushright}

\vspace{40pt}

\begin{center}

{\huge Kinetic equations for Bose-Einstein condensates from the 2PI effective action}

\end{center}

\vspace{1.5cm}

\begin{center}
R.~Baier $^a$, T.~Stockamp$^b$

\vspace{24pt}

Fakult\"at f\"ur Physik, Universit\"at Bielefeld,

D-33501 Bielefeld, Germany

\vspace{25pt}

\end{center}

\begin{abstract}

\noindent We use the 2PI effective action of a relativistic scalar field theory to derive
kinetic equations for a Bose-condensed system near the phase transition. We
start from equations of motions derived within a
$\frac{1}{N}$-expansion at NLO. In taking the
non-relativistic limit we obtain a generalized Gross-Pitaevskii equation for
the condensate field. Within the Popov approximation we explicitly compute the
collision term up to order $g^2$ using the Kadanoff-Baym formalism.
For the sake of self-consistency we derive in the same way a Boltzmann equation for the
non-condensate distribution function. The final results are in agreement to those
previously obtained by Griffin, Nikuni and Zaremba and by Stoof. 
\end{abstract}

\vspace{4cm}

\noindent $^amail:$ baier@physik.uni-bielefeld.de\\
$^bmail:$ stockamp@physik.uni-bielefeld.de

\newpage

\section{Introduction}
Since the first experimental realization of a  Bose condensed dilute
gas in 1995 there has been an intense effort to further understand this new
state of matter both from the experimental and theoretical point of view. The
most common theoretical ansatz for a wide range of questions is the famous
Gross-Pitaevskii equation of motion (e.o.m.) \cite{Dalfovo}. This equation for the Bose macroscopic
wavefunction $\Phi$ gives a complete description of the condensate at $T=0$. Near the phase transition one clearly has to take into account
the effects of the non-condensate atoms. One approach to this problem is to
derive a useful kinetic equation that includes the collisions between the
condensate and non-condensate part of the system. This was done by Griffin,
Nikuni and Zaremba \cite{zaremba} starting from a suitably generalized
Gross-Pitaevskii equation (see also the work by Imamovic-Tomasovic and Griffin
\cite{Griffin1, Griffin2, Griffin3}). Further approaches to a quantum kinetic
description of Bose condensed systems can, e.g., be found in \cite{Stoof, Gardiner:1998wk, Proukakis, Walser}.\\
In this paper we apply a method coming from non-equilibrium relativistic quantum field
theory. For a  recent
review of the so called two particle irreducible (2PI) effective action
techniques see \cite{Berges:2004yj}. This powerful method has
already been used in a variety of interesting issues like physics of
the early universe and heavy ion collisions
\cite{Berges:2002cz,Arrizabalaga:2004iw,Berges:2004ce}, and recently to the
non-equilibrium dynamics of Bose-Einstein condensates \cite{Rey}. We
use the e.o.m. for a N-component scalar field obtained within a
$\frac{1}{N}$-expansion of the 2PI effective action to next to leading
order (NLO) \cite{Aarts:2002dj}. We finally arrive at the same kinetic equation previously derived
in \cite{zaremba, Stoof}. 

\section{Field equation of motion}

\noindent Our starting point is the relativistic e.o.m. from the $\frac{1}{N}$-expansion of the 2PI effective
action at NLO. For a N-component real scalar field $\phi_a$ with mass $m$ and
$\frac{\lambda}{4!}\phi ^4$-interaction one has (see (27) in \cite{Aarts:2002dj})

\begin{equation}
-\left(\square + m^2 + \frac{\lambda}{6N} \left[\Phi_c(x) \Phi_c(x) +
 G_{cc}(x,x)\right]\right)\Phi_a(x) = K_a(x,x),
\end{equation}
with (see (35) in \cite{Aarts:2002dj})

\begin{eqnarray}
 &&K_a(x,y) = \frac{\lambda}{3N} \Phi_b(x)  G_{ba}(x,y)\nonumber \\ && \hspace{2cm} - i\frac{\lambda}{6N}
 \int_{\mathcal{C}}dz^0\int d^3z G_{bc}(x,z)  G_{bc}(x,z) K_a(z,y),
\end{eqnarray}
and the propagator $G$ given by

\begin{equation}
G_{ab}(x,y) = <T\phi_a(x)\phi_b(y)>-<\phi_a(x)><\phi_b(y)>.
\end{equation}
The mean field is denoted by $\Phi_a(x)=<\phi_a(x)>$.
In (3) $T$ means time ordering along the Schwinger-Keldysh
contour \cite{Schwinger:1960qe, Keldysh:1964ud, Bakshi:1962dv, Bakshi:1963bn}. The $z^0$-integral in (2) is taken along this contour as
well (see fig.1).\\ As we ultimately want to describe a Bose condensed gas we
have to take $N=2$, or $a=1,2$.\\ Next we perform the
non-relativistic limit by the following replacements

\begin{equation}
\Phi_a \longrightarrow \frac{1}{\sqrt{m}} \Phi_a^{NR},
\end{equation}

\begin{equation}
G \longrightarrow \frac{1}{m} G^{NR},
\end{equation}

\begin{equation}
\Pi \longrightarrow \frac{1}{m^2} \Pi^{NR},
\end{equation}

\begin{equation}
K_a \longrightarrow 2 \sqrt{m} K_a^{NR},
\end{equation}

\begin{equation}
\lambda \longrightarrow 12 m^2 g,
\end{equation}

\begin{equation}
(\square + m^2) \longrightarrow -2m (i\partial_{x^0} + \frac{\triangle}{2m}),
\end{equation}
where we define
\begin{equation}
\Pi(x,y) = -\frac{1}{2} G_{ab}(x,y) G_{ab}(x,y). 
\end{equation}
\noindent The dimensionful prefactors arise because of the different normalizations of
relativistic and non-relativistic fields. Throughout the paper we
set $\hbar=c=1$.\\
The non-relativistic e.o.m. now reads

\begin{eqnarray}
&&\left(i\partial_{x^0} + \frac{\triangle}{2m} - \frac{g}{2}\left[\Phi_c^{NR}(x)
\Phi_c^{NR}(x)+ G_{cc}^{NR}(x,x)\right]\right)\Phi_a^{NR}(x) =\nonumber\\&&
 g \Phi_b^{NR}(x) G_{ba}^{NR}(x,x) +2ig \int_{\mathcal{C}}dz^0\int d^3z
 \quad \Pi^{NR}(x,z) K_a^{NR}(z,x).
\end{eqnarray}
For a later comparison to other work on Bose condensed systems it is
convenient to express (11) in terms of one complex field $\Phi$. Thus we perform the following redefinition (we suppress from now
on the 'NR' index)

\begin{equation}
\phi_1  \longrightarrow \frac{1}{\sqrt{2}} (\Psi^{\ast} + \Psi), \qquad
\phi_2   \longrightarrow \frac{i}{\sqrt{2}} (\Psi^{\ast} - \Psi),
\end{equation}
with
\begin{equation}
\Psi = \Phi + \tilde{\Psi},
\end{equation}
and correspondingly for the expectation values

\begin{equation}
<\Psi> = \Phi , \qquad
<\tilde{\Psi}> = 0.
\end{equation}
The interpretations of $\Phi$ as the condensate and $\tilde{\Psi}$ as
non-condensate complex fields are obvious. The components of the propagator
matrix $G_{ab}$ are replaced by
\begin{equation}
G_{11}(x,x) \longrightarrow \frac{1}{2} (\tilde{m}^{\ast} + 2\tilde{n} + \tilde{m}),
\end{equation}

\begin{equation}
G_{22}(x,x) \longrightarrow -\frac{1}{2} (\tilde{m}^{\ast} - 2\tilde{n} + \tilde{m}),
\end{equation}

\begin{equation}
G_{12}(x,x) \longrightarrow \frac{i}{2} (\tilde{m}^{\ast} -  \tilde{m}),
\end{equation}

\begin{equation}
G_{21}(x,x) = G_{12}(x,x),
\end{equation}
where we introduced the non-equilibrium non-condensate densities
\begin{equation}
\tilde{n}(x) = <\tilde{\Psi}^{\ast}(x) \tilde{\Psi}(x)>, \qquad
\tilde{m}(x) = <\tilde{\Psi}(x) \tilde{\Psi}(x)>.
\end{equation}
The e.o.m. (11) reads in terms of the new fields

\begin{eqnarray}
&&\left(i\partial_{x^0} + \frac{\triangle}{2m} - g\left[n_c(x) +
2\tilde{n}(x)\right]\right)\Phi(x)-g\tilde{m}(x)\Phi^{\ast}(x) =  \nonumber\\&&
 2ig  \int_{\mathcal{C}}dz^0\int d^3z \Pi(x,z) K(z,x),
\end{eqnarray}
with
\begin{eqnarray}
K(x,y)&=& \frac{1}{\sqrt{2}} (K_1(x,y) + iK_2(x,y)) \nonumber\\
&=& g\Phi(x) <T\tilde{\Psi}^{\ast}(x) \tilde{\Psi}(y)>+ g\Phi^{\ast}(x)
<T\tilde{\Psi}(x) \tilde{\Psi}(y)> \nonumber\\&+& 2ig \int_{\mathcal{C}}dz^0\int d^3z \Pi(x,z) K(z,y),
\end{eqnarray}
and the non-equilibrium density in the condensate
$n_c(x) = \frac{1}{2}(\Phi_1^2(x) + \Phi_2^2(x)) = \left|\Phi(x)\right|^2$.\\
It is convenient to  write the r.h.s. of (20) in terms of $\Pi ^{\gtrless}$ and  $K ^{\gtrless}$. These are
defined as \cite{baym}
\begin{equation}
\Pi ^>(x,y) = \Pi (x,y) \qquad \text{for} \qquad  x^0 > y^0,
\end{equation}

\begin{equation}
\Pi ^<(x,y) = \Pi (x,y) \qquad \text{for} \qquad x^0 < y^0,
\end{equation}
and equivalently for  $K ^{\gtrless}$. \\
Using the relation

\begin{eqnarray}
 && \int_{\mathcal{C}}dz^0\int d^3z \quad \Pi(x,z) K(z,x)=\nonumber \\ &&
  \int_{t^0}^{x^0}dz^0\int d^3z \left(\Pi ^> (x,z) K ^< (z,x)- \Pi ^< (x,z) K ^> (z,x)\right)
\end{eqnarray}
finally leads to the e.o.m.

\begin{eqnarray}
&&(i\partial_{x^0} + \frac{\triangle}{2m} -g[n_c(x) +
2\tilde{n}(x)])\Phi(x) -g\tilde{m}(x)\Phi^{\ast}(x) = \nonumber \\ &&
2ig \int_{t^0}^{x^0}dz^0 \int d^3z (\Pi ^> (x,z) K ^< (z,x)- \Pi ^< (x,z) K ^>
(z,x)),
\end{eqnarray}
together with (21) for $K ^{\gtrless}$.\\
This equation can be seen as a generalization of the time dependent
Gross-Pitaevskii equation. In particular, the RHS contains interactions
between the condensate and the surrounding non-condensate cloud.

\begin{figure}[h]\begin{center}
\epsfig{file=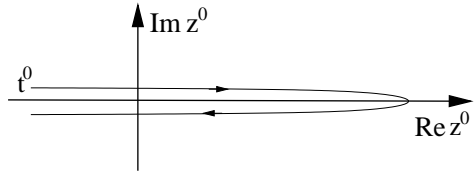,height=2.3cm}
\caption{Schwinger-Keldysh contour in the complex $z^0$-plane}
\end{center}\end{figure}

\subsection{Derivation of the kinetic equation}
Our goal in this section is to derive a kinetic equation for $\Phi$ in terms of the
non-condensate distribution functions. This equation will have a clear
physical interpretation. We have to make some
approximations as the very concept of a distribution function
$f(\vec{k},\vec{x},t)$ cannot be strictly applied to a quantum mechanical
problem. First of all, we will work within the Popov approximation \cite{Popov}, i.e. we neglect the
anomalous correlators $<\tilde{\Psi}\tilde{\Psi}>$ and
$<\tilde{\Psi}^{\ast}\tilde{\Psi}^{\ast}>$. In addition, we suppose the
coupling constant $g$ to be small as in current experiments the densities near
the phase transition are such that the gas is in the weak coupling limit
\cite{Dalfovo}.\\
One then obtains
\begin{equation}
\Pi ^{\gtrless}(x,y) \simeq - <\tilde{\Psi}(x) \tilde{\Psi}^{\ast}(y)>^{\gtrless}
<\tilde{\Psi}^{\ast}(x) \tilde{\Psi}(y)>^{\gtrless},
\end{equation}

\begin{equation}
K ^{\gtrless}(x,y) \simeq g \Phi (x) <\tilde{\Psi}^{\ast}(x) \tilde{\Psi}(y)>^{\gtrless},
\end{equation}
with $<\tilde{\Psi}(x) \tilde{\Psi}^{\ast}(y)>^< = <\tilde{\Psi}^{\ast}(y)
\tilde{\Psi}(x)>$ and so on. We take only the first term of the iterative equation (21) for
$K^{\gtrless}$ what leads to a collision term up to order $g^2$.\\
Following the well known approach of Kadanoff and Baym \cite{baym}, we now assume that
the correlation length of the system under consideration is sufficiently small and one can
write

\begin{equation}
<\tilde{\Psi}^{\ast}(y) \tilde{\Psi}(x)> = \int \frac{d^3p}{(2\pi)^3}
 e^{i\vec{p}\vec{s}-i\omega s^0} f(\vec{p},X),
\end{equation}

\begin{equation}
<\tilde{\Psi}(x) \tilde{\Psi}^{\ast}(y)> = \int \frac{d^3p}{(2\pi)^3}
 e^{i\vec{p}\vec{s}-i\omega s^0}\left(1 +  f(\vec{p},X)\right),
\end{equation}
where $s=x-y$ and $X=\frac{x+y}{2}$ are the relative and center of mass
coordinates, respectively, while $p=(\omega,\vec{p})$ denotes the
non-condensate quasiparticle four-momentum. The dependence of the correlation
functions (28) and (29) on the relative time $s^0=x^0-y^0$ is assumed to be
determined by these quasiparticles. According to (39), given later,  the
Hartree-Fock energy is given by $\omega_p=\frac{\vec{p}^2}{2m}+2g[n_c(x) +
\tilde{n}(x)]$ (see also \cite{zaremba}).\\
Inserting the above approximations in (25) gives
\begin{eqnarray}
&&\left(i\partial_{x^0}+\frac{\triangle}{2m}-g[n_c(x) +
2\tilde{n}(x)]\right)\Phi (x)= 2ig\Phi (x)\int_{t^0}^{x^0}dz^0 \int
d^3z\nonumber \\&& \times \int \prod_{i=1}^3\left(\frac{d^3p_i}{(2\pi)^3}\right) e^{i(m\vec{v}_c -\vec{p}_1
  -\vec{p}_2+\vec{p}_3) (\vec{z}-\vec{x}) -i(\epsilon_c -\omega_1 -\omega_2
  +\omega_3 ) (z^0 - x^0)}\nonumber \\
&&\times \bigl[ f_2 \left(1 + f_3\right) f_1 - \left(1 + f_2\right)
f_3 \left(1 +  f_1 \right)\bigr],
\end{eqnarray}
where $f_i$ means $f(\vec{p}_i,X)$. In writing (30) the expansion (following
\cite{zaremba})\\ $\Phi (z) \simeq \Phi (x) e^{im\vec{v}_c (\vec{z}-\vec{x}) - i{\epsilon}_c (z^0
  -x^0)}$ is applied, with $\vec{v}_c$ and $\epsilon_c$ being the condensates atoms local velocity and energy,
respectively.\\
In taking the limit $t^0 \longrightarrow -\infty$ one may approximate (again
following \cite{zaremba}) the time
integral in (30) by

\begin{equation}
\int_{-\infty}^{x^0} dz^0 e^{i({\epsilon}_c -\omega_1 -\omega_2
  +\omega_3  )(x^0
  -z^0)} f\left(\frac{x^0 + z^0}{2}\right) \simeq f(x^0)\pi\delta({\epsilon}_c -\omega_1 -\omega_2
  +\omega_3).
\end{equation}
The space dependence of the distribution functions may be simplified using a
gradient expansion around $\vec{x}$ giving
$f(\vec{X}) \simeq f(\vec{x})$.
The $\vec{z}$-integral in (30) can then easily be performed and one finally obtains

\begin{eqnarray}
&&\left(i\partial_{x^0} + \frac{\triangle}{2m} - g[n_c(x) +
2\tilde{n}(x)]\right)\Phi(x) =  \nonumber \\ && \frac{ig^2 \Phi (x)}{(2\pi)^5} \int 
d^3p_1 d^3p_2 d^3p_3 \quad \delta(m\vec{v}_c -\vec{p}_1
  -\vec{p}_2 +\vec{p}_3) \delta(\epsilon_c -\omega_1 -\omega_2
  +\omega_3)  \nonumber \\ &&\times\bigl[f_2(1 + f_3)f_1 - (1 + f_2)f_3(1 + f_1)\bigr].
\end{eqnarray}
This equation is in agreement with the result derived in \cite{zaremba}. The physical interpretation is obvious:
the first term in the integrand of the RHS corresponds to the gain term,
i.e. it describes the collisions of two non-condensate atoms scattering into a
non-condensate and a condensate state. Similarly, the second term describes
the inverse process leading to a loss of condensate states. The collision term
is shown pictorially in figure 2. Dashed lines correspond to the condensate and
full ones to the non-condensate field.

\begin{figure}[h]\begin{center}
\epsfig{file=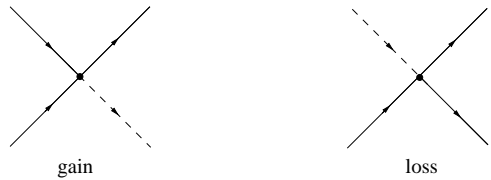,height=2.3cm}
\caption{pictorial representation of the collision integral}
\end{center}\end{figure}

\section{Propagator equation}

In the previous section we derived a kinetic equation for the condensate field $\Phi$
in terms of the non-condensate distribution function $f$. In order to obtain a
complete set of equations we thus have to find the Boltzmann equation
fulfilled by $f$. Our starting point is the propagator equation of motion
derived within the $\frac{1}{N}$-expansion of the 2PI effective action to NLO (see (36)
in \cite{Aarts:2002dj})

\begin{eqnarray}
&&-\left(\square + m^2 + \frac{\lambda}{6N} \left[\Phi_c(x) \Phi_c(x) +
 G_{cc}(x,x)\right]\right) G_{ab}(x,y) =
 i\delta_{ab}\delta_{\mathcal{C}}(x-y)\nonumber \\ &&+\Phi_a(x)K_b(x,y)
 -i \int_{\mathcal{C}}dz^0\int d^3z D(x,z)G_{ac}(x,z)G_{cb}(z,y),
\end{eqnarray}
with (see (34) in \cite{Aarts:2002dj})
\begin{eqnarray}
&&D(x,y)= i \frac{\lambda}{3N}\delta_{\mathcal{C}}(x-y)+
\frac{\lambda}{3N}K_a(y,x)\Phi_a(x) \nonumber \\ && -i \frac{\lambda}{6N}
\int_{\mathcal{C}}dz^0\int d^3z G_{ab}(x,z)G_{ab}(x,z)D(z,y).
\end{eqnarray}
We now sum equation (33) over suitable values of the indices
(a,b), in other words we take $(1,1)+(2,2)+i(1,2)-i(2,1)$. Next we perform the
non-relativistic limit and the field redefinition as before, leading to
\begin{eqnarray}
&&(-i\partial_{x^0} + \frac{\triangle}{2m} - gn(x))<T\tilde{\Psi}^{\ast}(x)
\tilde{\Psi}(y)> = \frac{i}{2}\delta_{\mathcal{C}}(x-y) +
\Phi^{\ast}(x)K(x,y)\nonumber \\ && - \frac{i}{2} \int_{\mathcal{C}}dz^0\int
d^3z D(x,z) \bigl[<T\tilde{\Psi}^{\ast}(x)
\tilde{\Psi}^{\ast}(z)><T\tilde{\Psi}(z)
\tilde{\Psi}(y)>\nonumber \\ &&\hspace{3.5cm} + <T\tilde{\Psi}^{\ast}(x)
\tilde{\Psi}(z)><T\tilde{\Psi}^{\ast}(z)
\tilde{\Psi}(y)>\bigr],
\end{eqnarray}
with
\begin{equation}
n(x)=n_c(x) +\tilde{n}(x),
\end{equation}
\begin{eqnarray}
&&D(x,y)=2ig\delta_{\mathcal{C}}(x-y) +4g\Phi^{\ast}(x)K(y,x)
+4g\Phi(x)K^{\ast}(y,x)\nonumber \\ && +2ig\int_{\mathcal{C}}dz^0\int
d^3z \Pi(x,z) D(z,y),
\end{eqnarray}
and $K$ given in (21).
Note that the correct non-relativistic replacement of the Klein-Gordon operator is here
\begin{equation}
(\square + m^2) \longrightarrow -2m (-i\partial_{x^0} + \frac{\triangle}{2m}),
\end{equation}
as it acts on $\tilde{\Psi}^{\ast}$.

\subsection{Boltzmann equation}
One may now derive a Boltzmann equation for the non-condensate
distribution function $f$. Starting from equation
(35) we apply again the Popov approximation and we neglect contributions in higher
order than $g^2$, giving 
\begin{eqnarray}
&&(-i\partial_{x^0} + \frac{\triangle}{2m} -2gn(x))<T\tilde{\Psi}^{\ast}(x) \tilde{\Psi}(y)>=\nonumber \\ &&
\frac{i}{2}\delta_{\mathcal{C}}(x-y)  -2ig^2 \int_{\mathcal{C}}dz^0\int
d^3z \\&& \times \bigl[     \Phi^{\ast}(x)\Phi(z)<T\tilde{\Psi}^{\ast}(x)
\tilde{\Psi}(z)><T\tilde{\Psi}(x)
\tilde{\Psi}^{\ast}(z)><T\tilde{\Psi}^{\ast}(z)
\tilde{\Psi}(y)>  \nonumber \\&&  + 
\Phi^{\ast}(z)\Phi(x)<T\tilde{\Psi}(z)
\tilde{\Psi}^{\ast}(x)><T\tilde{\Psi}^{\ast}(x)
\tilde{\Psi}(z)><T\tilde{\Psi}^{\ast}(z)
\tilde{\Psi}(y)>\nonumber \\&&  +
\Phi^{\ast}(x)\Phi(z)<T\tilde{\Psi}^{\ast}(z)
\tilde{\Psi}(x)><T\tilde{\Psi}^{\ast}(x)
\tilde{\Psi}(z)><T\tilde{\Psi}^{\ast}(z)
\tilde{\Psi}(y)>\nonumber \\&&  +
<T\tilde{\Psi}^{\ast}(x)
\tilde{\Psi}(z)><T\tilde{\Psi}(x)
\tilde{\Psi}^{\ast}(z)><T\tilde{\Psi}^{\ast}(x)
\tilde{\Psi}(z)><T\tilde{\Psi}^{\ast}(z)
\tilde{\Psi}(y)>
\bigr].\nonumber
\end{eqnarray}
Following, e.g., Blaizot and Iancu (see
section 2.3 of \cite{Blaizot:2001nr}) and using the techniques described in
the previous section we introduce the relative and center of mass coordinates $s$ and $X$ like
before and perform a Fourier transform according to (28) and (29).\\
As a result the Boltzmann equation finally becomes (with $T=X^0$)
\begin{equation}
\bigl[\partial_T+\frac{1}{m}\vec{p}\cdot\partial_{\vec{X}}-2g(\partial_{\vec{X}}n(X))\cdot\partial_{\vec{p}}\bigr]f(\vec{p},\vec{X},T)
= C_{12} + C_{22}.
\end{equation}
The r.h.s. of the above equation consists of two collision integrals describing
two body collisions between non-condensate atoms ($C_{22}$) and collisions
involving one condensate atom ($C_{12}$). They explicitly read
\begin{eqnarray}
&&C_{22}=\frac{2g^2}{(2\pi)^5}\int
d^3p_1d^3p_2d^3p_3\delta(\vec{p_1}+\vec{p_2}-\vec{p_3}-\vec{p})\delta(\omega_1+\omega_2-\omega_3-\omega)\nonumber \\&&
\hspace{2.6cm}\times\bigl[f_1 f_2 (1+f_3)(1+f)-ff_3 (1+f_2)(1+f_1)\bigr],
\end{eqnarray}
\begin{eqnarray}
&&C_{12}=\frac{4g^2n_c(X)}{(2\pi)^2}\int
d^3p_1d^3p_2\delta(m\vec{v_c}+\vec{p_1}-\vec{p_2}-\vec{p})\delta(\epsilon_c+\omega_1-\omega_2-\omega)\nonumber \\&&
\hspace{3.2cm}\times\bigl[f_1(1+f_2)(1+f)-ff_2(1+f_1)\bigr]\nonumber \\&&
\hspace{0.9cm} +\frac{2g^2n_c(X)}{(2\pi)^2}\int
d^3p_1d^3p_2\delta(m\vec{v_c}-\vec{p_1}-\vec{p_2}+\vec{p})\delta(\epsilon_c-\omega_1-\omega_2+\omega)\nonumber \\&&
\hspace{3.2cm}\times\bigl[f_1f_2(1+f)-f(1+f_2)(1+f_1)\bigr].
\end{eqnarray}
This result is in agreement with the one derived by Griffin, Nikuni and
Zaremba \cite{zaremba} and by Stoof \cite{Stoof}.\\
\\
The approach followed in this paper provides a systematic approximation scheme
for the non-equilibrium evolution of a Bose-Einstein condensate. Starting from
the nonperturbative equations (25,35) for scalar fields including scattering and
memory effects it gives as approximation a practicable system of kinetic
equations (32,40) at order $g^2$ describing the dynamical properties of condensate
formation. These equations have already been tested numerically using N-body
simulations \cite{Jackson}. \\
A full solution of the coupled 2PI equations (25,35), however, requires extensive
numerical work which is beyond the scope of this paper.

\section{Acknowledgements}
We would like to thank E.~Calzetta, B.~Mihaila and E.~Zaremba for useful
remarks. T.S. is supported by DFG.

\end{document}